\documentstyle[floats,eqsecnum,prd,aps,amsfonts,epsf]{revtex}
\begin{document}
\twocolumn[\hsize\textwidth\columnwidth\hsize\csname
@twocolumnfalse\endcsname
\title{Spectral Methods for Numerical Relativity.  The Initial Data
  Problem} 
\author{Lawrence E. Kidder}
\address{Center for Radiophysics and Space Research, Cornell
  University, Ithaca, New York, 14853}
\author{Lee Samuel Finn}
\address{Center for Gravitational Physics and Geometry, The
  Pennsylvania State University, University Park, Pennsylvania 16802}
\maketitle

\begin{abstract}
Numerical relativity has traditionally been pursued via finite
differencing. Here we explore pseudospectral collocation (PSC) as an
alternative to finite differencing, focusing particularly on the
solution of the Hamiltonian constraint --- an elliptic partial
differential equation --- for a black hole spacetime with angular
momentum and for a black hole spacetime superposed with gravitational
radiation.
In PSC, an approximate solution, generally expressed as a sum over a
set of orthogonal basis functions (e.g., Chebyshev polynomials), is
substituted into the exact system of equations and the residual
minimized.  For systems with analytic solutions the approximate
solutions converge upon the exact solution exponentially as the number
of basis functions is increased.  Consequently, PSC has a high
computational efficiency: for solutions of even modest accuracy we
find that PSC is substantially more efficient, as measured by either
execution time or memory required, than finite differencing;
furthermore, these savings increase rapidly with increasing accuracy.

For example, investigating the Hamiltonian constraint equation for a
black hole with angular momentum we find that, where finite difference
solutions require a resolution of $1024\times384$
(radial$\times$angular) grid points to find a solution of fractional
error $10^{-5}$ in the ADM mass, a PSC solution achieves the same
accuracy with only $12\times4$ collocation points. Furthermore, the
fractional error is reduced to $10^{-10}$ by increasing the PSC
resolution to $24\times8$, while the same increase in the finite
difference solution would require (if it were possible) an increase in
resolution by a factor of 300 in each dimension. Commensurate with the
doubling of the resolution in each of the two dimensions the computing
time required to find the spectral solution increase by a factor of
$2^2$ while the computing time required by the finite difference
method would increase by a factor of $300^2$, or $10^5$.

The solution provided by PSC is an analytic function given everywhere,
not just at the collocation points. Consequently, no interpolation
operators need to be defined to determine the function values at
intermediate points and no special arrangements need to be made to
evaluate the solution or its derivatives on the boundaries. 
Since the practice of numerical relativity by finite differencing has
been, and continues to be, hampered by both high computational
resource demands and the difficulty of formulating acceptable finite
difference alternatives to the analytic boundary conditions, PSC
should be further pursued as an alternative way of formulating the
computational problem of finding numerical solutions to the field
equations of general relativity.
\end{abstract}

\vskip2pc]

\section{Introduction and summary}

The partial differential equations (PDE) of numerical relativity have
typically been solved using finite difference methods.  In finite
differencing (FD) one first chooses a finite number of coordinate
``grid'' points $x_{n}$ and approximates the space and time
derivatives in the PDEs by ratios of differences between field and
coordinate values on the grid.  With a choice of grid and
``differencing scheme'' for converting derivatives to ratios of
differences, the equations of general relativity are approximated by a
system of algebraic equations whose solution approximates that of the
underlying PDEs.

In this paper we explore an alternative method for solving the elliptic
PDEs encountered in numerical relativity: pseudospectral collocation
(PSC).  In PSC one begins by postulating an approximate solution,
generally as a sum over some finite basis of polynomials or
trigonometric functions.  The coefficients in the sum are determined
by requiring that the residual error, obtained by substituting the
approximate solution into the exact PDEs, is minimized in some
suitable sense.  Thus, if one describes FD as finding the exact
solution to an approximate system of equations, one can describe PSC
as finding an approximate solution to the exact equations.

Pseudospectral collocation has been applied successfully to solve
problems in many fields, including fluid dynamics, meteorology,
seismology, and relativistic astrophysics (cf.\ 
\cite{boyd89a,canuto88a,fornberg96a,bonazzola98a}).  Its advantage
over FD arises for problems with smooth solutions, where the
approximate solution obtained using PSC converges on the actual
solution {\em exponentially} as the number of basis functions is
increased. The approximate FD solution, on the other hand, never
converges faster than algebraically with the number of grid points.
While the computational cost per ``degree of freedom'' --- basis
functions for PSC, grid points for FD --- is higher for PSC than for FD,
the computational cost of a high accuracy  
PSC solution is a small fraction of the cost of
an equivalent FD solution. Even for problems in which only
modest accuracy is needed, PSC generally results in a significant
computational savings in both memory and time compared to FD,
especially for multidimensional problems.

The elliptic equations of interest here are the constraint equations
that must be solved as part of the general relativistic Cauchy initial
data problem. We focus on two axisymmetric problems: the initial data
for a black hole spacetime with angular momentum, and a spacetime with
a black hole superposed with 
gravitational waves (Brill waves). Solutions for both of
these problems have been found by others using FD
\cite{choptuik86a,cook90a,bernstein94a}: our focus here is to
demonstrate the use of PSC for these problems and compare the
computational cost of high accuracy solutions obtained using both PSC
and FD.

In section~\ref{sec:init} we review briefly the key constraint
equations that arise in the traditional space-plus-time decomposition
of the Einstein field equations and describe three different elliptic
problems --- a nonlinear model problem whose analytic solution is
known, the nonlinear Hamiltonian constraint equation for an
axisymmetric black hole spacetime with angular momentum, and the
Hamiltonian constraint equation for a spacetime with a black hole
superposed with Brill waves --- that have been solved using FD and that we
solve here using PSC.  We describe PSC in section~\ref{sec:specmeth},
and compare the computational cost of PSC and FD for high accuracy
solutions in section~\ref{sec:compare}. In section~\ref{sec:specimpl}
we solve the problems described in section~\ref{sec:init} using PSC and
compare the performance of PSC with that obtained by other authors
using FD. In section~\ref{sec:concl}, we discuss our conclusions and their
implications for solving problems in numerical relativity. Finally,
whether by FD or PSC the solution of a nonlinear elliptic system
involves solving a potentially large system of (nonlinear) algebraic
equations. We describe the methods we use for solving them in appendix
\ref{sec:solvsys}.

\section{Initial Value Equations}
\label{sec:init}

\subsection{Introduction}

The general relativistic Cauchy initial value problem requires that we
specify the metric and extrinsic curvature of a three-dimensional
spacelike hypersurface.  These quantities cannot be specified
arbitrarily: rather they must satisfy a set of constraint equations,
which are a subset of the Einstein field equations.  Arnowitt, Deser,
and Misner (ADM) \cite{arnowitt62a} were the first to formulate the
Cauchy initial value problem in relativity in this way; however, the
most common expression of this $3+1$ decomposition is due to York
\cite{york79a}.

In the York formulation of the ADM equations the spacelike
hypersurfaces are taken to be level surfaces of some spacetime scalar
function $\tau$.  The generalized coordinates and conjugate momentum
are the three-metric $\gamma_{ij}$ induced on the spacelike
hypersurface by the space-time metric and the extrinsic curvature
${K}_{ij}$ of the spacelike hypersurface.\footnote{This approach is by
  no means unique.  For example, recent work by Choquet-Bruhat, York
  and collaborators (cf.\ \cite{anderson99a} and references therein)
  on a different choice of generalized coordinates and momenta have
  yielded approaches in which the evolution equations form a
  first-order symmetric and hyperbolic (FOSH) system. Many powerful
  numerical methods for solving FOSH systems exist and the numerical
  relativity community is only now beginning to explore how these
  solution techniques can be brought to bear on the field equations in
  this form.} Position on each surface is described by a set of
spatial coordinates $x_i$ (where Latin indices run from 1 to 3 and
indicate spatial coordinates on the slice), so that the line-element
on the hypersurface is
\begin{equation}
{}^{(3)}ds^2 = \gamma_{ij}dx^i\,dx^j.
\end{equation}

The normal $\bbox{n}$ to the spacelike
hypersurface is everywhere timelike. The time coordinate direction
$\bbox{t}$, however, need not be exactly along the normal. We can write 
$\bbox{t}$ in terms of $\bbox{n}$ and the spatial vectors that span 
the tangent space of the hypersurface:
\begin{equation}
\bbox{t} = \alpha \bbox{n} + \beta^i{\partial\over\partial x^i},
\end{equation}
where the {\em lapse\/} $\alpha$ describes how rapidly time elapses as
one moves along the hypersurface normal $\bbox{n}$, and the {\em
  shift\/} 
\begin{equation}
\vec\beta = \beta^i{\partial\over\partial x^i},
\end{equation}
is a vector field confined entirely to the hypersurface tangent space
that describes how the spatial coordinates are shifted, relative to
$\bbox{n}$, as one moves from one hypersurface to the next. The lapse
and shift are free functions: they correspond to the freedom to
specify the evolution of the coordinate system that labels points in
spacetime.

The {\em space-time\/} line element at any point on a hypersurface is
related to the spatial metric at that point, the lapse, and the shift:
\begin{eqnarray}
ds^2 &=& g_{\mu \nu} dx^\mu dx^\nu \nonumber \\ 
&=& - \left( \alpha^2 - \beta_a \beta^a 
\right) dt^2 + 2\beta_i dx^i dt + \gamma_{ij} dx^i dx^j
\end{eqnarray}
(where Greek indices run from 0 to 3 and include the time coordinate
$t$, which is sometimes referred to as $x_0$).

Given a spacetime foliation, choice of ``field variables''
$\gamma_{ij}$ and $K_{ij}$, and coordinates (embodied in the lapse and
shift), the field equations can be decomposed into four {\em
constraint equations,} which $\gamma_{ij}$ and $K_{ij}$ must satisfy
on each slice, and six {\em evolution equations,} which describe how
the three-metric and extrinsic curvature evolve from one slice to the
next.  In this paper we consider only the problem of consistent
specification of initial data: {\em i.e.,} the solution of the
constraint equations.

The four constraint equations (in vacuum) are 
\begin{mathletters}
\label{eq:constraints}
\begin{eqnarray}
{}^{(3)}R + K^2 - K_{ab}K^{ab} &=& 0, \\
{}^{(3)}\nabla_a \left( K^{ia} - K \gamma^{ia} \right) &=& 0,
\end{eqnarray}
\end{mathletters}
where ${}^{(3)}R$ is the Ricci scalar associated with $\gamma_{ij}$,
${}^{(3)}\nabla_a$ is the covariant derivative associated with
$\gamma_{ij}$, and
\begin{equation}
K:= K_{ab}\gamma^{ab},
\end{equation}
is the trace of the extrinsic curvature $K_{ij}$.  Note that, as 
befits constraints, equations \ref{eq:constraints}
involve only derivatives of $\gamma_{ij}$ and $K_{ij}$ in the tangent
space of the corresponding hypersurface. 

\subsection{Conformal imaging formalism}

Our goal is to determine a $\gamma_{ij}$ and a $K_{ij}$ that satisfy
the constraint equations and boundary conditions.  These are both
symmetric tensors on a three-dimensional hypersurface; consequently,
between the two there are twelve functions that must be specified. 
Equations \ref{eq:constraints} place only four
constraints on these twelve functions.  In order to solve the initial
value problem, the remaining components of $\gamma_{ij}$ and $K_{ij}$
must be specified. York \cite{york79a} has developed a
convenient formalism, referred to as {\em conformal imaging,} for
dividing the spatial metric and extrinsic curvature into constrained
and unconstrained parts, which we summarize in this subsection.

Associate $\gamma_{ij}$ with a {\em conformal background three-metric}
$\bar{\gamma}_{ij}$ through a {\em conformal factor\/} $\psi$:
\begin{equation}
\gamma_{ij} = \psi^4 \bar{\gamma}_{ij}.
\end{equation}
The extrinsic curvature ${K}^{ij}$ is split into its trace $K$ and its
trace-free part
\begin{equation}
A^{ij} = K^{ij} - {1\over3} \gamma^{ij} K.
\end{equation}
The trace $K$ is treated as a given scalar function which will be
specified.  The trace-free extrinsic curvature $\bar{A}^{ij}$ of
the conformal metric $\bar{\gamma}_{ij}$ can be expressed in terms of
$\psi$ and $A^{ij}$:
\begin{equation}
\bar{A}^{ij} = \psi^{10} A^{ij}.
\end{equation}

The constraint equations (eqs.~\ref{eq:constraints}) can also be
expressed in terms of the conformal background metric and its
trace-free extrinsic curvature:
\begin{mathletters}
\label{eq:cons}
\begin{eqnarray}
\bar{\nabla}^2 \psi - \frac{1}{8} \bar{R} \psi - \frac{1}{12} {K}^2 \psi^5
+ \frac{1}{8} \bar{A}_{ab} \bar{A}^{ab} \psi^{-7} &=& 0,
\label{eq:Hamcon}\\
\bar{\nabla}_a \bar{A}^{ia} - \frac{2}{3} \psi^6 \bar{\gamma}^{ia}
\bar{\nabla}_a {K} &=& 0,
\label{eq:momcon}
\end{eqnarray}
\end{mathletters}
where ${}^{(3)}\bar{\nabla}_i$ is the covariant derivative associated
with $\bar{\gamma}_{ij}$. 
Equation \ref{eq:Hamcon} is generally referred to as the Hamiltonian
constraint, while equations \ref{eq:momcon} 
are generally referred to as the momentum constraints.
Since the only derivatives of $\bar{A}_{ij}$ appear in the combination 
of a divergence, only the longitudinal part of $\bar{A}_{ij}$ is 
constrained by the momentum constraints. 

Now turn to the boundary conditions.  We are interested in problems
with a single black hole.  Let the initial hypersurface be
asymptotically flat, so that on the hypersurface far from the black
hole the curvature vanishes.  Describe the black hole by an
Einstein-Rosen bridge ({\em i.e.,} by two asymptotically flat
three-surfaces connected by a throat) and insist that the spacetime be
inversion symmetric through the throat.  These choices impose the
boundary conditions
\begin{mathletters}
\begin{eqnarray}
    \lim_{r\rightarrow\infty}\psi(r) &=& 1\qquad\text{asymptotic 
    flatness},\\
    \left[{\partial\psi\over\partial r} + {\psi \over 2a}\right]_{r=a} 
    &=& 0 \qquad\text{inversion symmetry},
\end{eqnarray}
\end{mathletters}
on $\psi$ where $r=a$ is the coordinate location of the throat.
    
We can now describe how to solve the constraint equations.  Let
${K}$ vanish on the initial hypersurface; then the Hamiltonian and
the momentum constraints (eqs.~\ref{eq:cons})
decouple.\footnote{Such a hypersurface is said to have vanishing mean
curvature, or be {\em maximally embedded.}} Pick a conformal
background metric $\bar\gamma_{ij}$ (which determines
${}^{(3)}\bar{\bbox{\nabla}}$) and transverse-traceless part
$\bar{A}_{TT}^{ij}$ of the conformal extrinsic curvature.  Solve the
momentum constraints (eqs.~\ref{eq:momcon}) for the
longitudinal part of the trace-free conformal extrinsic curvature
$\bar{A}_{ij}$.  Together with $\bar{A}_{TT}^{ij}$ the trace-free conformal
extrinsic curvature is thus fully determined and the Hamiltonian
constraint (eq.\ \ref{eq:Hamcon}) can be solved for the conformal
factor.  This determines the three-metric $\gamma_{ij}$ and its 
extrinsic curvature $K^{ij}$ and completes the specification of the 
initial data. 

\subsection{Three test problems}

\subsubsection{Black hole with angular momentum}\label{sec:bham} 

Focus first on the initial data corresponding to an axisymmetric black
hole spacetime with angular momentum.  This problem was first examined
analytically by \cite{bowen80a}, and has been explored numerically by
\cite{choptuik86a,cook90a}.

Choosing the conformal background metric to be flat ({\em i.e.,}
$\bar{\gamma}_{ij}=\delta_{ij}$), \cite{bowen80a} found
an analytic solution to the momentum constraints
(eqs.~\ref{eq:momcon}) that carries angular momentum and obeys the
isometry condition at the black hole throat.  Draw through any point a
sphere centered on the black hole throat, and let ${n}^i$ be the
outward-pointing unit vector to the sphere there.  Letting $J^i$ be
the angular momentum of the {\em physical\/} space, the Bowen-York
solution to the momentum constraints is
\begin{equation}
\bar{A}_{ij} = \frac{3}{{r}^3} \left[ \epsilon_{aib}
  J^b{n}^a{n}_j + \epsilon_{ajb}J^b{n}^a{n}_i \right].
\end{equation}
Corresponding to this solution is the Hamiltonian constraint
(eq.~\ref{eq:Hamcon}) for the conformal factor $\psi$,
\begin{equation}
\bar{\nabla}^2 \psi + \frac{9}{4} \frac{J^2 \sin^2 \theta}{r^6}
\psi^{-7} = 0,
\end{equation} 
together with its boundary conditions,
\begin{mathletters}
\label{eq:bc}
\begin{eqnarray}
\lim_{r\rightarrow\infty} \psi (r) &=& 1, \\
\left[ \frac{\partial \psi}{\partial r} + \frac{1}{2a} \psi
\right]_{r=a} &=& 0, \\
\left(\frac{\partial \psi}{\partial \theta} \right)_{\theta = 0,\pi} &=& 0,
\end{eqnarray}
\end{mathletters}
which result from asymptotic flatness, the isometry condition at
the throat (which is located at ${r}=a$), and axisymmetry respectively.  The
equation with boundary conditions for the conformal factor is
solved on the domain $a\leq r <\infty$. 
Once the conformal factor is determined, the geometry of the
initial slice is completely specified.

A useful diagnostic of an initial data slice is
to compute the total energy contained in the slice.
\'{O} Murchadha and York \cite{omurchadha74a} have examined the ADM
energy (cf.\ \cite{arnowitt62a}) in terms of the conformal 
decomposition formalism giving
\begin{equation}
E_{ADM} = \hat E_{ADM} - \frac{1}{2\pi} \oint_\infty \bar{\nabla}^j 
\psi d^2\bar{S_j},
\end{equation}
where $\hat E_{ADM}$ is the energy of the conformal metric.
Thus when the conformal metric is flat, the ADM energy reduces to
\begin{equation}
E_{ADM} = - \frac{1}{2\pi} \oint_\infty \bar{\nabla}^j \psi d^2\bar{S_j},
\label{eq:admE}
\end{equation}
{\em i.e.,} it is proportional to the integral of the normal component
of the gradient of the conformal factor about the sphere at infinity. 

\subsubsection{A model problem}\label{sec:model}

Bowen and York \cite{bowen80a} also describe a nonlinear ``model'' of
the Hamiltonian constraint equation that can be solved
exactly, which we utilize in section \ref{sec:specimpl} to test our
code.  The model equation is
\begin{equation}
\bar{\nabla}^2 \psi + \frac{3}{4} {P^2 \over r^4} \left( 1 -
  \frac{a^2}{r^2} \right)^2 \psi^{-7} = 0,
\label{eq:modham2}
\end{equation}
with $P$ a constant. Together with the boundary
conditions described above (eqs.~\ref{eq:bc}), equation~\ref{eq:modham2}
has the solution
\begin{mathletters}
\begin{equation}
\psi = \left[ 1 + \frac{2E}{r} + 6 \frac{a^2}{r^2} + \frac{2a^2E}{r^3}
+ \frac{a^4}{r^4} \right]^{1/4},
\end{equation}
where 
\begin{equation}
E = \left( P^2 + 4a^2 \right)^{1/2}.
\end{equation}
\end{mathletters}
If we evaluate equation \ref{eq:admE} for this solution, we find that it
has ADM energy $E$. 

\subsubsection{Black hole plus Brill wave}
\label{sec:bhbw}

The second physical problem upon which we demonstrate the use of
spectral methods for numerical relativity is that of a black hole
superposed with a Brill \cite{brill59a} wave, a problem studied using 
FD by \cite{bernstein94a}.

Following \cite{bernstein94a}, let the initial slice be a spacetime
isometry surface ({\em i.e.,} time symmetric); then, the extrinsic
curvature $K_{ij}$ vanishes and the momentum constraints
(eqs.~\ref{eq:momcon}) are trivially satisfied. 

To determine the conformal factor the Hamiltonian constraint
(eq.\ \ref{eq:Hamcon}) must be solved, which requires the
specification of a conformal background metric. Let the
line-element of the conformal background
metric have the form
\begin{equation}
d\bar{s}^2 = \left[ e^{2q} \left( dr^2 + r^2 d\theta^2 \right) + r^2
  \sin^2\theta d\phi^2 \right],
\end{equation}
where
\begin{mathletters}
\begin{eqnarray}
q &:=& A \sin^n\theta \left\{ \exp\left[ 
- \left( \frac{\eta + \eta_0}{\sigma} \right)^2 
\right] \right. \nonumber \\ &&+ \left. \exp \left[ - 
\left(\frac{\eta - \eta_0}{\sigma} \right)^2 
\right] \right\} \label{eq:brillq},\\ 
\eta &:=& \ln{r \over a},
\end{eqnarray}
\end{mathletters}
$n$ is an even integer, and $A$, $\eta_0$, and $\sigma$ are constant
parameters that describe the superposed Brill wave's amplitude,
position, and width, respectively.
With this choice the Hamiltonian constraint
equation becomes
\begin{eqnarray}
\frac{\partial^2 \psi}{\partial r^2} 
&+& \frac{2}{r}\frac{\partial \psi}{\partial r}
+ \frac{1}{r^2} \frac{\partial^2 \psi}{\partial \theta^2} 
+ \frac{\cot \theta}{r^2} \frac{\partial \psi}{\partial \theta} \nonumber \\
&+& \frac{\psi}{4} \left(\frac{\partial^2 q}{\partial r^2} 
+ \frac{1}{r} \frac{\partial q}{\partial r} 
+ \frac{1}{r^2} \frac{\partial^2 q}{\partial\theta^2} \right) = 0.
\end{eqnarray}

The boundary conditions of asymptotic
flatness and inversion symmetry are again given by equations
\ref{eq:bc}.  Furthermore, since the conformal metric $\bar
\gamma_{ij}$ has no ``$1/r$'' parts in its expansion at infinity, its
energy vanishes and the ADM energy for these slices is given
by equation~\ref{eq:admE}.

\section{Spectral Methods}\label{sec:specmeth}

\subsection{Introduction}

Consider an elliptic differential equation, specified by the operator
$L$ on the $d$-dimensional open, simply-connected domain ${\cal D}$,
with boundary conditions given by the operator $S$ on the boundary
$\partial{\cal D}$:
\begin{mathletters}
\begin{eqnarray}
L({u})(\bbox{x}) &=& f(\bbox{x}) \qquad \bbox{x}\in{\cal
  D}, \\ 
S({u})(\bbox{x}) &=& g(\bbox{x}) \qquad \bbox{x}\in\partial{\cal
  D}.
\end{eqnarray}
\end{mathletters}
There may be more than one boundary condition, in which case we can
index $S$ and $g$ over the boundary conditions. 

Approximate the solution $u(\bbox{x})$ to this system as a sum
over a sequence of {\em basis functions\/} $\phi_k(\bbox{x})$ on
${\cal D}+\partial\cal D$,
\begin{equation}
u_N(\bbox{x}) = \sum_{k=0}^{N-1} \tilde{u}_k\phi_k(\bbox{x}),
\end{equation}
where the $\tilde{u}_k$ are constant coefficients. Corresponding to
the approximate solution $u_N$ is a residual $R_N$ on $\cal D$ and
$r_N$ on $\partial{\cal D}$:
\begin{mathletters}
\begin{eqnarray}
R_N &=& L(u_N) - f \qquad \text{on $\cal D$},\\
r_N &=& S(u_N) - g \qquad \text{on $\partial\cal D$}.
\end{eqnarray}
\end{mathletters}
The residual vanishes everywhere for the exact solution $u$.  

In PSC we determine the coefficients $\tilde{u}_k$ by requiring 
that $u_N$ satisfies the differential equation and boundary conditions
{\em exactly\/} at a
fixed set of {\em collocation points\/} $x_n$:
{\em i.e.,} we require that 
\begin{mathletters}
\begin{eqnarray}
0 &=& L[u_N(x_n)] - f(x_n)\qquad\text{for $x_n$ in ${\cal D}$,}\\
0 &=& S[u_N(x_n)] - g(x_n)\qquad\text{for $x_n$ on $\partial{\cal D}$,}
\end{eqnarray}
\end{mathletters}
for all $n$.  When the expansion functions and collocation points are
chosen appropriately a numerical solution of these equations 
can be found very efficiently.  In the following
subsection we discuss choices of the expansion basis and collocation
points.

\subsection{Expansion basis and collocation points}

In PSC we require that the approximate solution $u_N$ satisfies the
differential equation and boundary conditions exactly at the $N$
collocation points $x_{n}$.  The basis $\phi_k$ should not constrain
the values of the approximation at the collocation points;
correspondingly, we can write the basis as a set of $N$ functions
$\phi_{k}(x)$ that satisfy a discrete orthogonality relationship on
the collocation points $x_{n}$:
\begin{equation}
    \sum_{n=0}^{N-1}\phi_{j}(x_{n})\phi^{*}_{k}(x_{n}) = 
    \nu^{2}_k \delta_{jk} ,
\end{equation}
where the $\nu_k$ are normalization constants.  Note that the basis
functions are inextricably linked with the collocation points.

It is sometimes the case that the basis  can be chosen so 
that the boundary conditions are automatically satisfied. For example, 
consider a one-dimensional problem on the interval 
\begin{equation}
   { \Bbb{I}} = [-1,1].
\end{equation}
If the boundary conditions are periodic then each element of the basis
\begin{mathletters}
    \label{eq:fourier}
\begin{equation}
    \phi_{k}(x) = 
        \exp\left[\pi i \left(x+1\right)k\right],
\end{equation}
satisfies the boundary conditions; correspondingly, the approximate
solution $u_N$ automatically satisfies the boundary conditions. If, in
addition, we choose the collocation points
\begin{equation}
    x_{n} = {2n\over N}-1,
\end{equation}
then the basis satisfies the discrete orthogonality relation 
\begin{equation}
\delta_{jk} = {1 \over N} \sum_{n=0}^{N-1}\phi_{j}(x_{n})\phi^{*}_{k}(x_{n}).
\end{equation}    
\end{mathletters}

In an arbitrary basis, or with arbitrarily chosen collocation points,
finding the $\tilde{u}_{k}$ from the $u_{N}(x_{n})$ requires the
solution of a general linear system of $N$ equations in $N$ unknowns,
which involves ${\cal O}(N^{3})$ operations.  For the basis and
collocation points given in equations \ref{eq:fourier} the
$\tilde{u}_{k}$ can be determined from the $u_{N}(x_{n})$ quickly and
efficiently via the Fast Fourier Transform in ${\cal O}(N\ln N)$
operations.

Arbitrary derivatives of the $u_{N}$ can also be computed quickly:
writing
\begin{mathletters}
\begin{equation}
{d^{p}u_{N}\over dx^{p}} = \sum_{k=0}^{N-1}\tilde{u}_{k}^{(p)}\phi_{k}(x),
\end{equation}
we see immediately that
\begin{equation}
    \tilde{u}_{k}^{(p)} = \left(\pi i k\right)^{p}.
\end{equation}
\end{mathletters}
Consequently, any derivative of $u_{N}$ can be evaluated at all the
collocation points in just ${\cal O}(N\ln N)$ operations.

The ability to evaluate efficiently the derivatives of $u_{N}$ at the
collocation points is much more important than finding a basis whose
individual members satisfy the boundary conditions.  In the case of
periodic boundary conditions we can have our cake and eat it, too. 
More generally we choose a basis in which we can efficiently
compute the derivatives of $u_{N}$ at the collocation points and
require separately that the approximate solution $u_{N}$ satisfy the
boundary conditions at collocation points on the boundary.  

For general boundary conditions a basis of Chebyshev polynomials often
meets all of our requirements.\footnote{The geometry of a problem
  might suggest other expansion functions, such as Legendre polynomials; 
  however, a Chebyshev expansion does quite
  well and has the added convenience that, with appropriately chosen
  collocation points, only ${\cal O}(N\ln  N)$ are required to convert
  from the expansion coefficients to the function values at the
  collocation points and vice versa \cite{orszag80a}.}  Recall that the Chebyshev
polynomials are defined on $\Bbb{I}$ by
\begin{equation}
T_k(x) = \cos\left(k\cos^{-1} x\right).\label{def:cheb}
\end{equation}
A simple recursion relation allows us to find the derivative of $u_N$
as another sum over Chebyshev polynomials: if \footnote{For Chebyshev
  bases the conventional notation is that $k$ runs from $0$ to $N$,
  not $N-1$; thus, there are $N+1$ coefficients and collocation points.} 
\begin{equation}
u_N(x) = \sum_{k=0}^{N} \tilde{u}_k T_k(x),
\end{equation}
then
\begin{equation}
{du_N\over dx}(x) = \sum_{k=0}^{N-1} \tilde{u}'_k T_k(x),
\end{equation}
where
\begin{equation}
c_k \tilde{u}'_{k} = \tilde{u}'_{k+2}+2(k+1)\tilde{u}_{k+1},
\end{equation}
with
\begin{equation}
c_k = \left\{ \begin{array}{ll}
2 & \; \text{$k = 0$}\\
1 & \; \text{$k \geq 1$}.
\end{array}\right.
\label{eq:c}
\end{equation}

If we choose collocation points $x_n$ (for $0\leq n \leq N$) according to 
\begin{equation}
x_n = \cos {\pi n\over N},
\end{equation}
then the Chebyshev polynomials satisfy the discrete orthogonality
relation
\begin{equation}
  \delta_{jk} = {2\over N \bar{c}_k}
  \sum_{n = 0}^{N}
  {1\over\bar{c}_{n}} T_{j}(x_{n})T_{k}(x_{n}),
\end{equation}
where
\begin{equation}
\bar{c}_k = \left\{ \begin{array}{ll}
2 & \; \text{$k = 0$ or $N$} \\
1 & \; \text{otherwise}.
\end{array}\right.
\label{eq:cbar}
\end{equation}
Finally, exploiting the relation between the Chebyshev polynomials
and the Fourier
basis (cf.\ \ref{def:cheb}) allows us to find the $\tilde{u}_k$ from
the $u_N(x_n)$ in ${\cal O}(N\ln N)$ time using a fast transform
\cite[appendix B]{canuto88a}.  With an expansion basis of Chebyshev
polynomials and an appropriate choice of collocation points we can thus
evaluate derivatives of arbitrary order at the collocation points in
${\cal O}(N\ln N)$ operations.

For problems on an arbitrary domain of dimension $d$ greater than unity it
is rarely the case that we can find a basis which permits rapid
evaluation of derivatives. If the domain can be mapped smoothly to
${\Bbb{I}}^d$ then we can write
\begin{mathletters}
\begin{equation}
    u_{N^{(1)}\cdots N^{(d)}}(\bbox{x}) = 
    \sum_{k_{1}=0}^{N^{(1)}}\cdots
    \sum_{k_{d}=0}^{N^{(d)}} \tilde{u}_{k_{1}\cdots k_{d}}
    \phi_{k_{1}\cdots k_{d}}(\bbox{x}),
\end{equation}
where
\begin{eqnarray}
\phi_{k_{1}\cdots k_{d}}(\bbox{x}) &=& 
\prod_{\ell=1}^d\phi^{(\ell)}_{k_{\ell}}(x^{(\ell)}),\\
\bbox{x} &=& \left(x^{(1)},\ldots,x^{(d)}\right),
\end{eqnarray}
and the $\phi^{(\ell)}_k$, for fixed $\ell$, is a basis on $\Bbb{I}$ which
permits fast evaluation of derivatives with respect to its argument
({\em e.g.,} Chebyshev polynomials). Associated with each set of basis
functions are the collocation points $x^{(\ell)}_{n}$;
correspondingly, the collocation points associated with
$\phi_{k_1\cdots k_d}$ are just the $N_{1}\cdots N_{d}$-tuples
\begin{equation}
\bbox{x}_{n_1\cdots n_d} = \left(
x^{(1)}_{n_1},\cdots,x^{(d)}_{n_d}
\right).
\end{equation}
\end{mathletters}
With this choice of basis and collocation points we can evaluate
efficiently arbitrary derivatives of an approximation
$u_{N^{(1)}\cdots N^{(d)}}$.  If the domain cannot be mapped smoothly
to a $d$-cube in ${\Bbb{R}}^d$, either more sophisticated methods such
as domain decomposition \cite{boyd89a,canuto88a} must be used, or the
problem may not be amenable to solution by PSC.

\subsection{Solving the system of equations}

The expansion basis, collocation points and differential equation with
boundary conditions determine a system of equations for the
coefficients $\tilde{u}_k$ or, equivalently, the approximate solution
$u_N$ evaluated at the collocation points.  
Iterative solution methods (which require as few as ${\cal
  O}(N\ln  N)$ operations) work well to solve the kind of systems
of equations that arise from the application of a PSC method.

If the elliptic system being solved is linear then the algebraic
equations arising from either a FD or a PSC method are also linear and
a unique solution is guaranteed. If, on the other hand, the
differential system is nonlinear, then the equations arising from FD
or PSC are also nonlinear and a unique solution is not guaranteed.
Newton's method \cite[sec.\ 12.13 and appendices C and D]{boyd89a},
where one solves the linearized equations beginning with a guess and
then iterating, works well for these types of equations. As long as a
good initial guess is chosen, the iteration will usually converge.
In appendix \ref{sec:solvsys} we describe in detail the variant of
Newton's method (Richardson's iteration) that we have used to solve the
nonlinear system of algebraic equations that arise when we apply PSC
to solve the Hamiltonian constraint equations as posed in section
\ref{sec:init}.

\section{Comparing Finite Difference and Pseudospectral Collocation
  Methods} 
\label{sec:compare} 

\subsection{Introduction}

Finite differencing and pseudospectral collocation are alternative
ways to find approximate solutions to a system of differential
equations. Consider the Poisson problem in one dimension:
\begin{mathletters}
\label{eq:system}
\begin{equation}
  {d^2u\over dx^2} = f(x),\label{eq:poisson}
\end{equation}
on the interval $\Bbb I$ with Dirichlet boundary conditions
\begin{equation}
u(-1) = u(1) = 0. \label{eq:dbc}
\end{equation}
\end{mathletters}
In a FD approach to this problem we seek the values of $u$ at discrete
points $x_n$, say
\begin{equation}
x_n = -1 + {2n \over N},
\end{equation}
for $n = 0,1,\dots N$.  Algebraic equations
are found by approximating the differential operator
$d^2u/dx^2$ in equation \ref{eq:poisson} by a ratio of differences: {\em e.g.,}
\begin{equation}
{d^2u\over dx^2}(x_n) \simeq 
{u_{n+1}-2u_n+u_{n-1}\over\Delta x^2},
\end{equation}
for integer $n = 1,2,\dots N-1$
where
\begin{mathletters}
\begin{eqnarray}
u_n &:=& u(x_n),\\
\Delta x &:=& 2/N.
\end{eqnarray}
\end{mathletters}
With this discretization the differential equation \ref{eq:poisson}
yields $N-1$ equations for the $N+1$ unknown $u_n$. The boundary
conditions (eq.\ \ref{eq:dbc}) yield two more equations,
completely determining the $u_n$:
\begin{mathletters}
\begin{equation}
{u_{n+1}-2u_n+u_{n-1}\over\Delta x^2} = f(x_n) \qquad
\text{for
  $n = 1,2,\dots N-1$},
\end{equation}
\begin{eqnarray}
u_{0} &=& 0,\\
u_{N} &=& 0.
\end{eqnarray}
\end{mathletters}
The solution to these equations is the FD approximation to
$u(x)$ at the points $x_n$. 

The FD solution to equations \ref{eq:system} begins by approximating
the differential equations.  In the PSC method, on the other hand, we
first approximate the solution at all points in $\Bbb I$ by a sum over
a finite set of basis functions. For this example, we choose a Chebyshev
basis; so, we write
\begin{equation}
u_{N}(x) = \sum_{k=0}^{N} \tilde{u}_k T_k(x).
\end{equation}

Now insist that $u_{N}$ satisfies the differential equation and
boundary conditions exactly at the collocation points 
\begin{equation}
x_n = \cos{\pi n\over N},
\end{equation}
for $n = 0,1,\dots N$.
In particular, we require that the boundary conditions are satisfied
and that, in addition, the differential equation is satisfied
for integer $n$ ranging from $1$ to $N-1$:
\begin{mathletters}
\begin{eqnarray}
u_N(x_0) &=& 0,\\
u_N(x_N) &=& 0,\\
{d^2u_N \over dx^2}(x_n) &=& f(x_n).\label{eq:difeq}
\end{eqnarray}
\end{mathletters}
To evaluate equation \ref{eq:difeq} note that $d^2u_N/dx^2$ can be 
written as
\begin{mathletters}
\label{eq:2ndDeriv}
\begin{equation}
{d^2u_N \over dx^2}(x_n) = \sum_{m=0}^{N} d^{(2)}_{nm} u_N(x_m).
\end{equation}
The $d^{(2)}_{nm}$ can be determined by noting that 
\begin{equation}
{d^2u_N\over dx^2}(x_n) = \sum_{k=0}^{N} \tilde{u}_k'' T_k(x_n),
\end{equation}
with
\begin{eqnarray}
c_k \tilde{u}_k'' &=& \tilde{u}_{k+2}'' + 2(k+1) \tilde{u}_{k+1}',\nonumber\\
c_k \tilde{u}_k' &=& \tilde{u}_{k+2}' + 2(k+1) \tilde{u}_{k+1},
\end{eqnarray}
and
\begin{equation}
  \tilde{u}_k = {2\over N \bar{c}_k}
  \sum_{n = 0}^{N}
  {1\over\bar{c}_{n}} u_N(x_{n})T_{k}(x_{n}),
  \label{eq:speccoefs}
\end{equation}
\end{mathletters}
where $c_k$ and $\bar{c}_k$ are given by equations \ref{eq:c} and
\ref{eq:cbar}, respectively.

The result is, again, a set of algebraic equations for $u_N(x_n)$: the
values of the approximate solution at the collocation points.  Finding
the $u_N(x_n)$ yields an approximate solution to the differential
equation over the entire domain $\Bbb I$ since the spectral
coefficients $\tilde{u}_k$ are given by equation
\ref{eq:speccoefs}.\footnote{Alternatively, we could have constructed
a system of equations in terms of the unknown spectral coefficients. 
This would correspond to a spectral tau method: {\em cf.}
\cite{boyd89a,canuto88a}.}

For the linear problem posed here the solution to the
algebraic system of equations that arise in either a FD or PSC solution can
be solved directly or by any of the many standard iterative methods.
For nonlinear problems the systems are generally solved by linearizing
the equations about an initial guess and then iterating the solution
until it converges. We discuss one method of solution in appendix
\ref{sec:solvsys}.

\subsection{Convergence of approximations}

In either a FD or PSC solution to a differential equation with
boundary conditions we expect that, as $N$ tends to infinity, the
approximate solution should become arbitrarily accurate. For large
$N$, the $L_2$ error in a FD approximation converges upon the exact
solution as $N^{-p}$ for positive integer $p$. The value of $p$
depends on the smoothness of $f$ and the error in the approximation of
the differential operator (in the example above, $d^2/dx^2$). Assuming
that $f$ is smooth the rate of convergence (measured by the $L_2$
error of the FD solution) is $N^{-p}$ when the truncation error of
the differential operator is ${\cal O}(\Delta x^p)$.

In contrast, when the solution $u$ is smooth the error made by a
properly formulated spectral approximation decreases faster than any
fixed power of $N$ (where $N$ is now the number of collocation points
or basis functions).\footnote{In addition the individual spectral
  coefficient $\tilde u_k$ should decrease exponentially with $N$ once
  the problem is sufficiently resolved.} For a heuristic understanding
of this rapid convergence, note first that a PSC solution's
derivatives at each collocation point involve all the $\{ u_N(x_n) \}$
(cf.~eq.~\ref{eq:2ndDeriv}). Correspondingly, it is as exact as
possible, given the information available at the $N$ collocation
points.  This suggests that an order $N$ collocation spectral
approximation to the derivatives of the unknown should make errors on
order ${\cal O}(\Delta x^{N})$. The interval $\Delta x$, however, is
also proportional to $N^{-1}$; so, we expect that the error in the
spectral solution $u_N$ should vary as ${\cal O}(N^{-N})$. A more
rigorous analysis using convergence theory \cite[chapter 2]{boyd89a}
shows that for any function which is analytic on the domain of
interest, a Chebyshev expansion will converge exponentially ({\em
  i.e.} as ${\cal O}(e^{-N})$).  If the function is also periodic then a
Fourier expansion will converge exponentially.

\subsection{Computational cost of solutions}

The computational cost, in time, of a FD solution to a system of
elliptic differential equations scales linearly with the number of
grid points $N$ while the accuracy $\epsilon$ of the solution scales
as $N^{-p}$, where $p$ is the order of the FD operator
truncation error.  Correspondingly, the cost $K_{\text{FD}}$ for a
given accuracy 
scales as 
\begin{mathletters}
\begin{equation}
K_{\text{FD}} \sim \epsilon^{-1/p}.
\end{equation}
The cost $K_{\text{PSC}}$ of a PSC solution to the same system, on the
other hand, scales as $N\ln N$ (for an iterative solution) while
$\epsilon$ scales as $\exp(-N)$; consequently, the cost scales with
accuracy $\epsilon$ as
\begin{equation}
K_{\text{PSC}} \sim
-(\ln \epsilon)\,\ln \left(-\ln \epsilon\right).
\end{equation}
\end{mathletters}
Since it is the computational cost required to achieve a given
accuracy that is important, the more rapid convergence of a PSC
solution confers upon it a clear advantage.
This advantage is made clear by considering how the ratio of costs
scales with accuracy:
\begin{equation}
{K_{\text{PSC}}\over K_{\text{FD}}} \sim
-\epsilon^{1/p}\ln\epsilon\,\ln\left(-\ln\epsilon\right),
\end{equation} 
which tends to zero with $\epsilon$; consequently, increasing accuracy
with a PSC solution is always more efficient than with a FD
solution. 

The equations that arise from either a FD or PSC treatment of an
elliptic differential system are typically solved using iterative
methods; thus, {\em at fixed resolution\/} the storage requirements
for either solution method are equivalent. As we have seen, however,
fixed resolution does not correspond to fixed solution accuracy. As
the desired solution accuracy increases the storage requirements of a
PSC solution fall relative to those of an FD solution a factor of
$-\epsilon^{1/p}\ln\epsilon$.

\section{Solving the Hamiltonian Constraint}
\label{sec:specimpl}

\subsection{Nonlinear model problem}

As a first example we solve the model Hamiltonian constraint
equation described in section \ref{sec:model}, equation
\ref{eq:modham2}:
\begin{mathletters}
\begin{equation}
\bar{\nabla}^2 \psi + \frac{3}{4} {P^2 \over r^4} \left( 1 -
  \frac{a^2}{r^2} \right)^2 \psi^{-7} = 0,
\label{eq:modham5}
\end{equation}
for $r\in[a,\infty)$ with $P$ a constant and with the boundary conditions
\begin{eqnarray}
    \lim_{r\rightarrow\infty}\psi(r) &=& 1\qquad\text{asymptotic 
    flatness},\\
    \left[{\partial\psi\over\partial r} + \frac{1}{2a} \psi \right]_{r=a}
     &=& 0\qquad\text{inversion symmetry},\\
    \left(\frac{\partial \psi}{\partial \theta} \right)_{\theta = 0,\pi}
     &=& 0\qquad\text{axisymmetry}.\label{eq:angbc}
\end{eqnarray}
\end{mathletters}
For this model problem the solution $\psi$ is 
\begin{mathletters}
\label{eq:modsol}
\begin{equation}
\psi = \left[ 1 + \frac{2E}{r} + 6 \frac{a^2}{r^2} + \frac{2a^2E}{r^3}
+ \frac{a^4}{r^4} \right]^{1/4},
\end{equation}
where 
\begin{equation}
E = \left( P^2 + 4a^2 \right)^{1/2},
\end{equation}
\end{mathletters}
is also the ADM energy for the initial data (cf. eq.~\ref{eq:admE}). 

As described this problem is spherically symmetric; nevertheless, we
treat it as axisymmetric to illustrate the methods used to solve the
Hamiltonian constraint for the black hole with angular momentum (cf.\ 
sec.\ \ref{sec:bham}) and the black hole plus Brill wave problems (cf.\ 
sec.\ \ref{sec:bhbw}).

As a first step we map the domain $r\in[a,\infty)$, $\theta\in[0,\pi]$
to a square in ${\Bbb R}^{2}$: letting
\begin{mathletters}
\label{eq:mapxy}
\begin{eqnarray}
x &=& \frac{2a}{r} - 1,\\
y &=& \cos \theta,
\end{eqnarray}
\end{mathletters}
we have $x\in(-1,1]$ and $y\in[-1,1]$. 
In terms of the $(x,y)$ coordinates, the model Hamiltonian constraint
(eq.\ \ref{eq:modham5}) becomes
\begin{eqnarray}
\left( x+1 \right)^2
\frac{\partial^2 \psi}{\partial x^2} +&& \left( 1-y^2 \right) 
\frac{\partial^2 \psi}{\partial y^2} - 2y \frac{\partial \psi}{\partial y}
\nonumber \\ + \frac{3}{256} \left( \frac{P}{a} \right)^2 &&
\left( x+1 \right)^2\left( 3-2x-x^2 \right)^2 \psi^{-7} = 0,
\end{eqnarray}
subject to the boundary conditions
\begin{mathletters}
\label{eq:bcxy}
\begin{eqnarray}
\lim_{x\rightarrow-1}\psi &=& 1,\\
\left[{\partial \psi \over \partial x} - {1 \over 4} \psi \right]_{x=1} &=& 0.
\end{eqnarray}
\end{mathletters}
Note that with our choice of variables and expansion bases the angular
boundary conditions (eq.~\ref{eq:angbc}) are automatically satisfied.

Since $\psi$ is not periodic in either $x$ or $y$, we adopt a
Chebyshev basis for the approximate solution:
\begin{mathletters}
\begin{equation}
\psi_{N_x,N_Y} (x,y) = \sum_{j=0}^{N_x} \sum_{k=0}^{N_y} \tilde
\psi_{jk} T_j(x)T_k(y),\label{eq:expansion}
\end{equation}
with the corresponding collocation points
\begin{eqnarray}
    x_{j} &=& \cos{\pi j\over N_x},\\
    y_{k} &=& \cos{\pi k\over N_y}.
\end{eqnarray}
\end{mathletters}
For this problem, focus on approximations
\begin{equation}
\Psi_{\ell} = \psi_{4\ell,4},
\end{equation}
for integer $\ell$. We keep $N_y$ fixed as the model problem is
independent of $y$.

Following the discussion in appendix \ref{sec:solvsys}, solve the PSC
equations using Richardson's iteration with a second-order FD
preconditioner.  To obtain $\Psi_\ell$, we need an initial guess
$\Psi_\ell^{(0)}$ to begin the iteration.  For the lowest resolution
expansion ($N_x=4$) begin the iteration with the guess
\begin{equation}
\Psi^{(0)}_1(x,y) = \frac{(3+x)}{2},
\end{equation}
which is the trivial solution for $P=0$.  Applying Richardson's
iteration will then give us the approximate solution $\Psi_1$.
Through the expansion \ref{eq:expansion} this determines an
approximation for $\psi$ everywhere; in particular, it determines an
approximation at the collocation points corresponding to $N_x=8$,
which we then use as the initial guess for determining the approximate
solution $\Psi_{2}$. In this same way we use a lower-resolution
approximate solution as the initial guess for the approximate solution
at the next higher-resolution, {\em i.e.}
\begin{equation}
\Psi_\ell^{(0)} = \Psi_{\ell -1}.
\end{equation}

To investigate the accuracy of our solution as a
function of resolution (basis dimension for PSC, number of grid points
for FD) we evaluate a number of solutions differing only in resolution
and evaluate several different error measures.
\begin{enumerate}
\item For this problem we know the exact solution (cf.\ 
  \ref{eq:modsol}); so, we calculate the $L_2$ norm of the
  absolute error as a function of $\ell$:
  \begin{eqnarray}
    \Delta\Psi_{\ell} &=& \Biggl\{ \sum_{j=0}^{N_x}\sum_{k=0}^{N_y}
    \frac{1}{N_x N_y \bar{c}_j \bar{c}_k} \left[\Psi_\ell(x_j,x_k) 
	\right. \Biggr. \nonumber \\ && \Biggl. \left. -
        \psi(x_j,x_k)\right]^2\Biggr\}^{1/2}\nonumber\\
    &:=& \left|\left| \Psi_\ell - \psi \right|\right|_2,
    \label{eq:Delta}
  \end{eqnarray}
  where $c_k$ is given by equation \ref{eq:cbar}.
\item We can also characterize the convergence of the approximate
  solutions $\Psi_{\ell}$ by calculating the $L_2$ norm of the
  difference between the successive approximate solutions:
  \begin{equation}
    \delta\Psi_{\ell} = \left|\left| \Psi_\ell - \Psi_{\ell-1}
    \right|\right|_2.\label{eq:delta}
  \end{equation}
  The errors $\delta\Psi$ and $\Delta\Psi$ are defined for either FD or
  PSC solutions. 
\item We also evaluate, by analogy with $\Delta\Psi$ and $\delta\Psi$,
  the quantities
  \begin{eqnarray}
  \Delta{E}_{\ell} &=& | E_\ell - E |,\\
  \delta{E}_{\ell} &=& | E_\ell - E_{\ell-1} |,
  \end{eqnarray}
  where $E_\ell$
  is the ADM mass-energy associated with the approximate conformal
  factor $\Psi_\ell$. We evaluate $E$ using equation \ref{eq:admE}.
\item For PSC solutions only we define the relative error
  measure 
  \begin{equation}
    \delta\tilde{\Psi}_{\ell} = \sum_{j=0}^{N_x} \sum_{k=0}^{N_y} 
    \left|\tilde{\psi}^{(\ell)}_{jk} -
      \tilde{\psi}^{(\ell-1)}_{jk}\right|,
  \end{equation}
  which characterizes the changes in the spectral coefficients as the
  order of the approximation increases.
\end{enumerate}
For a properly formulated spectral method, all of our error measures should
decrease exponentially with $N$ if the solution to the problem is analytic.

\begin{figure}
\begin{center}
\epsffile{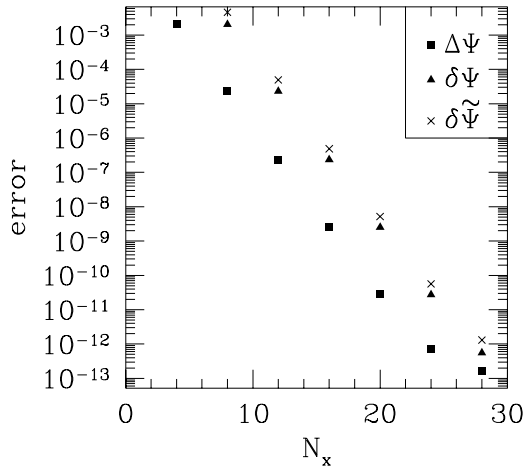}
\end{center}
\caption{Spectral convergence for a nonlinear model problem.  Plotted are
a measure of the absolute error $\Delta\Psi_{\ell}$, and two approximate 
measures of the error $\delta\Psi_{\ell}$ and $\delta\tilde{\Psi}_{\ell}$
as a function of $N_x$, the number of radial functions, for the case $P=1$.}
\label{fig:modprob}
\end{figure}

Figure \ref{fig:modprob} shows the absolute and relative errors
$\Delta\Psi_{\ell}$ and $\delta\Psi_{\ell}$, along with the change in
the spectral coefficients $\delta\tilde{\Psi}_{\ell}$, for $P=1$.
The exponential convergence of the solution with increasing $N_x$
is apparent.  Experience shows that as the problem becomes more
nonlinear ({\em i.e.,} $P$ becomes larger) more terms are needed in
the expansion in order to achieve the same accuracy.

\epsfxsize\columnwidth

This system of equations has also been solved using FD methods
\cite{cook90a}.  A point comparison is telling: in \cite{cook90a} a
second order accurate FD solution with a resolution of 1024 radial
points points were required for a solution with a
$\Delta{E}\simeq10^{-5}$, independent of $P$.  The PSC solution
described here achieves the same accuracy using an expansion with only
12 radial functions for $P=1$, and 24 functions for $P=10$.  In either
case a PSC solution with an accuracy of $\Delta E \approx 10^{-10}$ is
obtained by doubling the number of radial functions. To achieve the
same accuracy the FD approximation would require (assuming second
order FD) a resolution of $3\times10^5$ radial points.

\subsection{Black hole with angular momentum}

\epsfxsize\columnwidth
\begin{figure}
\begin{center}
\epsffile{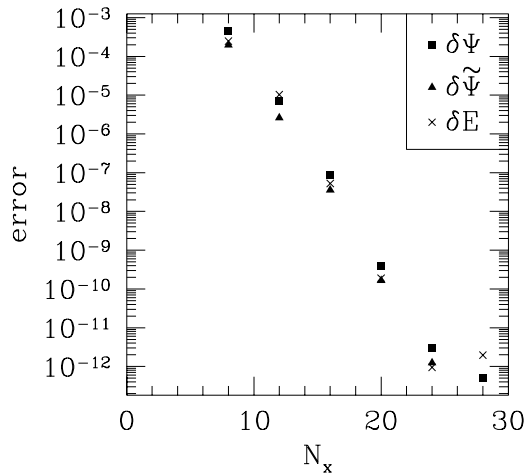}
\end{center}
\caption{Spectral convergence for the solution of the Hamiltonian constraint
equation for a black hole with angular momentum .  Plotted are three
approximate measures of the
error $\delta\Psi_{\ell}$, $\delta\tilde{\Psi}_{\ell}$ and $\delta E$ 
as a function of $N_x$, the number of
radial functions, for $J=1$.}
\label{fig:bhangmomj1}
\end{figure}

Now turn to consider a truly non-radial, but still axisymmetric,
problem: a rotating black hole (cf.\ \ref{sec:bham}).  As before (cf.\ 
\ref{eq:mapxy}) we map the semi-infinite domain $r\geq a$ to the
finite box $x\in(-1,1]$, $y\in[-1,1]$, obtaining the system of
equations
\begin{eqnarray}
\left( x+1 \right)^2
\frac{\partial^2 \psi}{\partial x^2} +&& \left( 1-y^2 \right) 
\frac{\partial^2 \psi}{\partial y^2} - 2y \frac{\partial \psi}{\partial y}
\nonumber \\ 
+ \frac{9}{64} \left( \frac{J}{a^2} \right)^2 &&\left( x+1 \right)^4
\left( 1-y^2 \right) \psi^{-7} = 0,
\end{eqnarray}
subject to the boundary conditions given in equations \ref{eq:bcxy}. 

\begin{figure}
\begin{center}
\epsffile{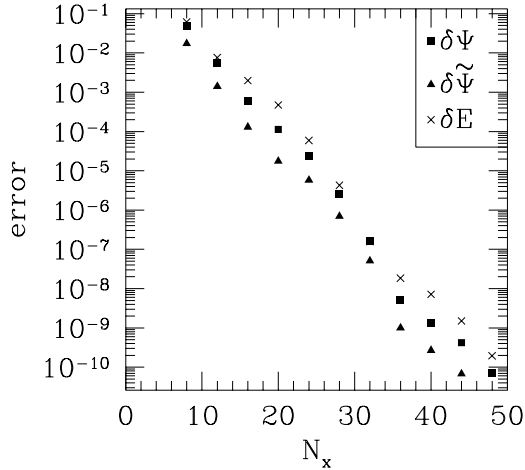}
\end{center}
\caption{Same as figure \ref{fig:bhangmomj1} with $J=100$.}
\label{fig:bhangmomj100}
\end{figure}

For this problem we do not have the exact solution; so, we consider
only the relative errors $\delta\Psi$, $\delta\tilde{\Psi}$ and
$\delta E$. Figure \ref{fig:bhangmomj1} (\ref{fig:bhangmomj100}) shows
these quantities as functions of $N_x$ for $J/M^2$ equal to 1 (100).
For these solutions $\Psi_{\ell} = \psi_{4\ell,N_y}$, where initially
$N_y=4$ and is incremented by two\footnote{Along with axisymmetry, this
problem has equatorial plane symmetry so $\Psi_{\ell}$ is even in $y$.
By exploiting this symmetry, we could reduce our number of angular
functions by a factor of two} whenever the difference between 
$\delta \Psi_\ell$ with and without the increment was greater than
ten percent.
Again we see rapid, exponential convergence
of the solution with $N$.

This problem has also been solved using second order FD
\cite{cook90a}. For a solution accuracy $\delta{E}\simeq10^{-5}$,
\cite{cook90a} found that a resolution 1024 radial and 384 angular
grid points was required, roughly independent of the value of $J$. We
find that PSC achieves the same accuracy with an expansion basis of 12
radial (and 4 angular) functions for $J=1$, and 24 radial (and 8
angular) functions for $J=100$. Solution accuracies of $10^{-10}$ can
be obtained for the PSC solution simply by doubling the size of the
expansion basis (in $x$ and $y$). For a similar increase in accuracy
of the FD solution a grid approximately 300 times larger in each
dimension would be required.

\epsfxsize\columnwidth

\subsection{Black hole plus Brill wave}

\epsfxsize\columnwidth
\begin{figure}
\begin{center}
\epsffile{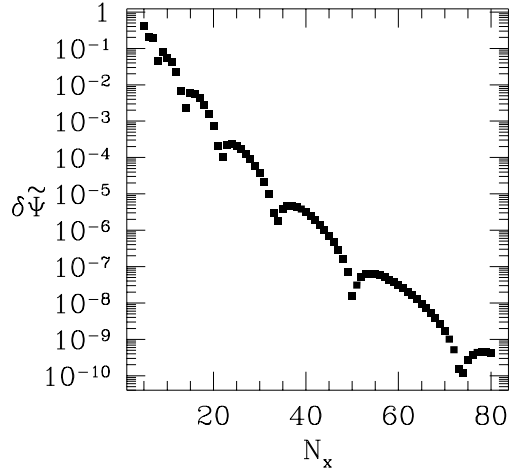}
\end{center}
\caption{Spectral convergence for the solution of the Hamiltonian constraint
equation for a black hole plus Brill wave.  Plotted is an
approximate measure of the
error $\delta\tilde{\Psi}_{\ell}$ as a function of $N_x$, the number of
radial functions, for the case $A=\eta_0=\sigma=1$, $n=2$.}
\label{fig:bhbrill}
\end{figure}

As a final example we consider the Hamiltonian constraint for a black
hole superposed with a Brill wave. After mapping this problem to the
$(x,y)$ domain we obtain the system of equations
\begin{mathletters}
\begin{equation}
\left( x+1 \right)^2
\frac{\partial^2 \Psi}{\partial x^2} + \left( 1-y^2 \right) 
\frac{\partial^2 \Psi}{\partial y^2} - 2y \frac{\partial \Psi}{\partial y}
 + \frac{\Psi R}{4} = 0,
\end{equation}
with 
\begin{equation}
R = \left( x+1 \right)^2
\frac{\partial^2 q}{\partial x^2} + (x+1) \frac{\partial q}{\partial x}
+ \left( 1-y^2 \right)\frac{\partial^2 q}{\partial y^2}
- y \frac{\partial q}{\partial y},\label{eq:R}
\end{equation}
\end{mathletters}
where $q$ is given by equation \ref{eq:brillq}, and subject to the boundary
conditions \ref{eq:bcxy}.

\begin{figure}
\begin{center}
\epsffile{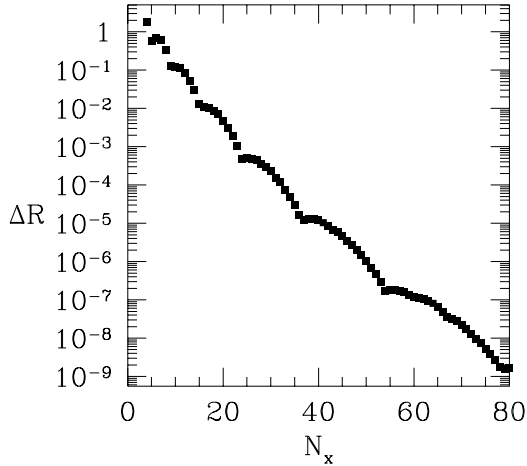}
\end{center}
\caption{The error in the spectral representation of $R$ (equation 
\ref{eq:R}) for the case shown in figure \ref{fig:bhbrill}}
\label{fig:ricci}
\end{figure}

In figure \ref{fig:bhbrill} we show $\delta\tilde\Psi$ as a function
of $N_x$ for the Brill wave parameters $\sigma = A = \eta_0 = 1$ and
$n = 2$.  For these solutions $\Psi_{\ell} = \psi_{4\ell,N_y}$ where initially
$N_y=4$, and is incremented by two whenever the difference between 
$\delta \Psi_\ell$ with and without the increment was greater than
ten percent.
The convergence, while rapid, is not
quite exponential. In addition, the nearly exponentially decreasing
error is impressed with a wave that is nearly periodic in spectral
resolution $\log N_x$. We attribute this behavior to the resolution of
the factor $R$ (cf.\ eq.\ \ref{eq:R} and also eq.\ \ref{eq:brillq} for
$q$).  Figure \ref{fig:ricci} shows the the error $\Delta R$ obtained when
we form approximate $R_{N_x,N_y}$ according to
\begin{mathletters}
\begin{equation}
R_{N_x,N_y} = \sum_{j=0}^{N_x}\sum_{k=0}^{N_y} \tilde{R}_{jk}
T_j(x)T_j(y),
\end{equation}
with	
\begin{equation}
\tilde{R}_{jk} = 
{4\over N_xN_y \bar{c}_k\bar{c}_j}
\sum_{\ell=0}^{N_x}\sum_{m=0}^{N_y} {1 \over \bar{c}_\ell\bar{c}_m}
T_{j}(x_{\ell})T_{k}(y_{m})R(x_\ell,y_m).
\end{equation}
\end{mathletters}
The structure in the
solution is the same as the structure in the Chebyshev approximation
to $R$.

\epsfxsize\columnwidth

This problem has also been solved using FD methods
\cite{bernstein94a}, enabling us to compare the resolution required
for approximate FD or PSC solution for a given accuracy.  With second
order FD a solution whose error $\delta E$ is $3 \times 10^{-5}$ required a
resolution of 400 radial and 105 angular grid points.  To achieve the
same accuracy the PSC solution described here requires a basis of only
36 radial (and 12 angular) Chebyshev polynomials.

\section{Discussion}
\label{sec:concl}
Pseudospectral collocation is a very efficient way of solving the
nonlinear elliptic equations that arise in numerical relativity.
These problems typically have smooth solutions; correspondingly, the
approximate solutions obtained using pseudospectral collocation
converge upon the exact solution exponentially with the number of
collocation points.  As a result, the cost of a high accuracy
pseudospectral solution is not significantly greater than the cost of
a similar solution of moderate accuracy.  Since the computational
burden of solving the pseudospectral collocation equations with a
given number of collocation points is no greater than that required to
solve the finite difference equations for the same number of grid
points, the computational demands of a pseudospectral collocation
solution are far less than those of a finite difference solution for
even modest accuracy.

Another important advantage of a pseudospectral collocation solution
over a finite differencing solution involves the formulation of the
boundary conditions. In a finite difference solution the boundary
conditions must be reformulated as finite difference equations
or incorporated approximately into the formulation of the finite
difference operator of the differential equations being solved. This
generally involves the introduction of auxiliary boundary conditions,
which are not part of the original problem. For example, consider the
second order elliptic equation on $\Bbb I$:
\begin{mathletters}
    \label{eq:fdexample}
\begin{eqnarray}
    {d^{2}u\over dx^{2}} &=& f(x),\label{eq:fdexamplea}\\
    u(-1) &=& u(1) = 0.
\end{eqnarray}
\end{mathletters}
A fourth-order accurate finite difference approximation to the
differential operator $d^{2}/dx^{2}$ is
\begin{equation}
    {16(u_{j-1}-2u_{j}+u_{j+1}) - (u_{j-2}-2u_{j}+u_{j+2})\over\Delta
    x^{2}} = f(x_{j}),
\end{equation}
where
\begin{equation}
    u_{j} = u(j\Delta x).
\end{equation}
Before this finite difference operator can be used in equation
\ref{eq:fdexample} it must be modified at the grid points
$-1+\Delta x$ and $1-\Delta x$ since $-1-\Delta x$ and $1+\Delta x$
both lie outside the computational domain. In this case, four boundary
conditions are required (at $x$ equal to $-1$, $-1+\Delta x$,
$1-\Delta x$ and $1$) even though the second order equation
\ref{eq:fdexamplea} properly admits of only two boundary conditions.

In pseudospectral collocation, on the other hand, no auxiliary
boundary conditions need be formulated: the approximate solution is
expressed as an analytic function, which is required to satisfy the
boundary condition equations exactly at the collocation points on the
boundary.

These advantages of pseudospectral collocation solution come at a
cost.  When properly implemented the computational expense of
pseudospectral collocation may be considerably less than the expense
of finite differencing; however, the difficulty of implementation is
greater.  The efficient solution of the algebraic equations arising
from pseudospectral collocation generally require the use of
sophisticated iterative methods.  Additionally, the exact solution
itself should be smooth on the computational domain in order that
exponential convergence is attained.  Finally, and perhaps most
importantly, for problems of dimension greater than unity the
computational domain must be sufficiently simple that it can be mapped
to ${\Bbb I}^{d}$ or be decomposed into sub-domains that
can each be mapped to ${\Bbb I}^{d}$ ({\em e.g.,} an L-shaped region
can be decomposed into two regions, each of which can be mapped to
${\Bbb I}^{2}$).

We have not here investigated the application of pseudospectral
collocation techniques to evolution problems.  Pseudospectral
collocation methods have been used to solve problems in other fields
({\em e.g.,} fluid dynamics) with great success
\cite{canuto88a,fletcher84a}. Our own experience in applying these
techniques to evolution problems in numerical relativity shows
promise, but is not yet complete.

\begin{acknowledgements}

It is a pleasure to acknowledge the support of the National Science
Foundation (PHY/ASC93-18152, ARPA supplemented, PHY 99-00111, PHY
99-96213, PHY 95-03084, PHY 99-00672, PHY 94-08378).  L.S.F. would also
like to acknowledge the support of the Alfred P. Sloan Foundation.

\end{acknowledgements}


\appendix

\section{Solving the pseudospectral collocation equations}
\label{sec:solvsys}

In this appendix we describe one method of solving the nonlinear
equations that arise from applying a PSC method to a nonlinear elliptic
system of equations
\begin{mathletters}
\label{eq:systema}
\begin{eqnarray}
L(u) &=& f\qquad\text{on $\cal D$},\\
S(u) &=& g\qquad\text{on $\partial\cal D$}.
\end{eqnarray}
\end{mathletters}

Choosing an expansion basis and corresponding collocation points, the
PSC solution of these equations is fully characterized by
the values of $u_N$ at the collocation points $x_n$: from these the
coefficients of the expansion and all the derivatives of the
approximate solution can be determined. Write the values of the
approximate solution $u_N$ at the collocation points $x_n$ as a vector
$\bbox{U}$, 
\begin{equation}
{\bbox{U}}_n = u_N(x_n).
\end{equation}
Corresponding to the approximate solution $u_N$ is a {\em residual\/}
$R_N$ on $\cal D$ and $r_N$ on $\partial{\cal D}$:
\begin{mathletters}
\begin{eqnarray}
R_N &=& L(u_N) - f \qquad \text{on $\cal D$},\\
r_N &=& S(u_N) - g \qquad \text{on $\partial\cal D$}.
\end{eqnarray}
\end{mathletters}
The residual vanishes everywhere for the exact solution $u$.
Write the values of the
residual at the collocation points $x_n$ as a vector
$\bbox{R}$, 
\begin{equation}
{\bbox{R}}_n = \left\{ \begin{array}{ll}
R_N(x_n) & \; \text{$x_n$ on $\cal D$} \\
r_N(x_n) & \; \text{$x_n$ on $\partial\cal D$}.
\end{array} \right.
\end{equation}
The PSC solution $\bbox{U}$ satisfies the algebraic equations
\begin{equation}
{\bbox{R}}\left[{\bbox{U}}\right] = 0.
\label{eq:reszero}
\end{equation}

Before describing how to solve equation \ref{eq:reszero} for a
nonlinear system ({\em i.e.,} nonlinear $L$ or $S$) we describe the
method of solution for a linear system.

When the system of differential equations \ref{eq:systema} is linear
so is the system of algebraic  
equations \ref{eq:reszero}. In this case we can write 
\begin{equation}
\bbox{\Lambda}\bbox{U} = \bbox{F},\label{eq:lin}
\end{equation}
where $\bbox{\Lambda}$ is a matrix and $\bbox{F}$ is a vector whose
components take on the values of $f$ and $g$ evaluated at the
collocation points in the domain ${\cal D}$ and its boundary
$\partial{\cal D}$.  In PSC the matrix $\bbox{\Lambda}$ is typically
full.  Direct solution methods require ${\cal O}(N^3)$ operations for
such systems; for efficiency such systems are generally solved by
iterative methods, which typically requires many fewer operations to
find an accurate solution.

A simple and effective iterative method for solving equation
\ref{eq:lin} is Richardson's iteration. 
Suppose we have a guess
$\bbox{V}^{(i)}$ to the solution $\bbox{U}$ of equation
\ref{eq:lin}. A better approximation to $\bbox{U}$ is
$\bbox{V}^{(i+1)}$ given by
\begin{equation}
\bbox{V}^{(i+1)} = \bbox{V}^{(i)} - \omega \bbox{R}^{(i)},
\end{equation}
where the residual $\bbox{R}^{(i)}$ vector is given by
\begin{equation}
\bbox{R}^{(i)}=\bbox{\Lambda}\bbox{V}^{(i)}-\bbox{F},
\end{equation}
and $\omega$ is a relaxation parameter, which must be determined. The
optimal value of $\omega$ and the rate of convergence of the
iterations depend upon the eigenvalues of $\bbox{\Lambda}$.  For
Richardson's iteration the optimal $\omega$ is
\begin{equation}
\omega_{\text{opt}} = {2 \over\lambda_{\max}+\lambda_{\min}},
\end{equation}
where $\lambda_{\max}$ and $\lambda_{\min}$ are the largest and
smallest eigenvalues of $\bbox{\Lambda}$. 
This choice minimizes the
spectral radius $\rho$,
\begin{equation}
\rho = {\lambda_{\max} - \lambda_{\min} \over \lambda_{\max} +
  \lambda_{\min}},
\end{equation}
of the iteration matrix, $G = I - \omega \Lambda$.
The convergence rate of the iteration is
\cite{canuto88a}
\begin{equation}
{\cal R} = -\ln\rho.
\end{equation}
The reciprocal of $\cal R$ measures the number of iterations
required to reduce the error by a factor of $e$.

Richardson's iteration is, by itself, not necessarily more efficient
than a direct solution method. Consider, for example, the 
second-order differential equation
\begin{mathletters}
\label{eq:examplea}
\begin{eqnarray}
{d^2 u\over dx^2} &=& f(x) \qquad x\in(-1,1),\\
u(-1) &=& u(1) = 0
\end{eqnarray}
\end{mathletters}
(cf.\ also section \ref{sec:compare}).  A PSC solution with a
Chebyshev expansion basis leads to an operator $\bbox{\Lambda}$ with a
spectral condition number $\lambda_{\max}/\lambda_{\min}$ that is
${\cal O}(N^2)$. This gives a rate of convergence ${\cal R}\sim{\cal
  O}(N^{-2})$; correspondingly, ${\cal O}(N^2)$ iterations are
required to obtain a reasonable solution. Since each iteration
requires ${\cal O}(N \ln N)$ operations ({\em i.e.,} it is
asymptotically dominated by the cost of evaluating the derivatives
$d^2u/dx^2$ given the $N+1$ $u_N(x_n)$) the total cost of obtaining a
solution $\bbox{U}$ is ${\cal O}(N^{3} \ln N)$, which is slightly {\em
  more\/} expensive than a direct solution.

We can speed the convergence of Richardson's iteration by solving an
equivalent problem whose spectral condition number is better behaved. 
Introduce the {\em preconditioning\/} matrix $\bbox{H}$ and consider
the equivalent system
\begin{equation}
\bbox{H}^{-1}\bbox{\Lambda}\bbox{U}=\bbox{H}^{-1}\bbox{F}.
\end{equation}
Now given an approximation $\bbox{V}^{(i)}$ to $\bbox{U}$, a better
approximation $\bbox{V}^{(i+1)}$ is given by 
\begin{equation}
\bbox{V}^{(i+1)} = \bbox{V}^{(i)} - \omega'
\bbox{H}^{-1}\bbox{R}^{(i)},
\end{equation}
where $\bbox{R}^{(i)}$ is given as before and $\omega'$ is related to the
eigenvalues of the linear operator
$\bbox{H}^{-1}\bbox{\Lambda}$. 

In practice we never actually invert the preconditioning
matrix $H$; instead we solve
\begin{equation}
\bbox{H}\left(\bbox{V}^{(i+1)} - \bbox{V}^{(i)}\right) = -
\omega'\bbox{R}^{(i)},
\label{eq:pcsys}
\end{equation}
for successive approximations. In order that this equation for
successive approximations should converge rapidly we require
a preconditioning matrix $\bbox{H}$ such that
\begin{itemize}
\item equation \ref{eq:pcsys} is inexpensive to solve, and 
\item the spectral condition number $\kappa'$ of  $\bbox{H}^{-1}\bbox{\Lambda}$
  is close to unity.
\end{itemize}
If $\bbox{H}^{-1}$ is a good approximation of
$\bbox{\Lambda}^{-1}$ then the second condition will be
satisfied; consequently, we look for approximations to
$\bbox{\Lambda}$ for which equation \ref{eq:pcsys} is inexpensive to
solve.

The operator $\bbox{\Lambda}$ arises from a system of differential
equations. For one-dimensional problems a low-order FD
approximation to this operator (with grid points coincident with the
collocation points) gives rise to a banded system with a small number
of bands close to the main diagonal. When this FD
operator is used as the preconditioner the system of equations
\ref{eq:pcsys} can be solved efficiently using direct
methods.\footnote{For more details on the use of FD
  operators as preconditioners for spectral problems see
  \cite{orszag80a}.} 

For instance, in the example considered here (eq.~\ref{eq:examplea})
we can set $\bbox{H}$ to
be the second-order accurate FD operator corresponding
to $L$. The eigenvalues of the preconditioned operator
$\bbox{H}^{-1}\bbox{\Lambda}$ are all in the range $1 \leq \lambda^{PC}_p
\leq \pi^2/4$: {\em i.e.,} the spectral condition number is
independent of $N$.  In this case the optimal relaxation parameter
is 
\begin{equation}
\omega'_{\text{opt}} \approx \frac{4}{7},
\end{equation}
and each iteration reduces the residual by a factor of approximately
$7/3$ (independent of $N$) \cite{orszag80a}.  The asymptotic cost of
finding a solution is thus proportional to the cost of evaluating the
residual, ${\cal O}(N\ln N)$, which is much more rapid than solution
via a direct method or Richardson's iteration without a
preconditioner.

For higher dimensional problems the FD preconditioner still leads to a
banded system with a small number of non-zero bands; however, some of
those bands are found far from the main diagonal and equation
\ref{eq:pcsys} can no longer be solved efficiently using direct
methods. If $N$ becomes so large that the cost of solving these
equations with the FD preconditioner is too great, then the equations
for the successive approximations can themselves be solved
iteratively, other preconditioners can be explored (cf.\ 
\cite{boyd89a,canuto88a,zang84a}), or the original equations can be
solved using another iterative technique, such as multigrid
\cite{choptuik86a}. For the problems considered in this paper $N$
never became so large that a direct solution of equation
\ref{eq:pcsys} with the FD preconditioner was problematic.

If the equations \ref{eq:systema} are nonlinear, the algebraic
equations satisfied by $\bbox{U}$ are similarly nonlinear. Write the
nonlinear equations as
\begin{equation}
\bbox{\cal L}(\bbox{U}) = \bbox{F},
\end{equation}
where $\bbox{\cal L}$ is a nonlinear function of $\bbox{U}$.  In order
to solve this nonlinear system of equations, we apply Newton's
iteration \cite[sec.\ 12.13 and appendices C and D]{boyd89a}.  For
equation \ref{eq:reszero}, Newton's iteration is
\begin{equation}
{\bbox{\Lambda}}_{\bbox{V}^{(i)}}(\bbox{V}^{(i+1)}-\bbox{V}^{(i)}) 
= - \bbox{R}^{(i)},\label{eq:Rn}
\end{equation}
where ${\bbox{\Lambda}}_{\bbox{V}^{(i)}}$ is the {\em linear\/} operator
that arises from linearizing $\bbox{\Lambda}$ about $\bbox{V}^{(i)}$
and $\bbox{R}^{(i)}$ is the {\em nonlinear} residual given by
\begin{equation}
\bbox{R}^{(i)} = \bbox{\cal L}(\bbox{U}) - \bbox{F}.
\end{equation}
Equation \ref{eq:Rn} is a linear system to be solved at each step
of Newton's iteration. In the same way as before we can introduce a
preconditioner, in which case we have the 
nonlinear Richardson's iteration
\begin{equation}
\bbox{H}\left(\bbox{V}^{(i+1)} - \bbox{V}^{(i)}\right) = -
\omega'\bbox{R}^{(i)}.
\end{equation}
Here $\bbox{H}$ is any suitable preconditioning matrix for
${\bbox{\Lambda}}_{\bbox{V}^{(i)}}$.  For the problems solved in
section \ref{sec:specimpl} we used as a preconditioning matrix a
second-order accurate FD operator corresponding to the derivative
terms of $\cal L$ {\em ignoring} the nonlinear terms. (Equivalently we
could have used the FD operator corresponding to the linearized
operator, but for the problems we examined this was not necessary.)

As a quick demonstration, consider the example problem
\begin{eqnarray}
\label{eq:tabexamp}
{d^2u\over dx^2} -  e^x [ (1-\pi^2) \sin (\pi x) + 2\pi \cos(\pi
x) ] &=& 0,\\
u(-1) = u(1) &=& 0.
\end{eqnarray}
We have evaluated approximate solutions to this problem using a
second-order accurate FD approximation and a PSC approximation on a
Chebyshev basis. For this problem we know the exact solution,
\begin{equation}
u(x) = e^x \sin(\pi x).
\end{equation}
Table \ref{tab:examprob} lists $\Delta u_{PSC}$ and $\Delta u_{FD}$
(cf.\ \ref{eq:Delta}) for increasing $N$ (number of
grid points for the FD approximation, basis dimension for the PSC
approximation).  The rapid convergence of PSC is apparent.  The
second-order FD solution requires 128 points to equal the moderate
accuracy of an eighth-order PSC solution.  In order to match the high
accuracy of the 16th-order PSC solution would require a second-order
FD solution with 6.5 $\times 10^{4}$ points.

\begin{table}
\caption{The values of the absolute error $\Delta u_{PSC}$ of a
PSC calculation, as well as the absolute 
error of a second-order FD calculation $\Delta u_{FD}$ for several 
values of $N$ for the example problem (equation \ref{eq:tabexamp}).
For $N>16$, the PSC solution is contaminated with roundoff errors.}
\label{tab:examprob}
\begin{tabular}{d|dd}
$N$&$\Delta u_{PSC}$&$\Delta u_{FD}$\\
\tableline\\[-9pt]
4&1.7 $\times 10^{-1}$& 2.5 $\times 10^{-1}$ \\
8&3.2 $\times 10^{-4}$& 5.4 $\times 10^{-2}$ \\
16&6.9 $\times 10^{-10}$& 1.3 $\times 10^{-2}$ \\
32&-& 3.3 $\times 10^{-3}$ \\
64&-& 8.2 $\times 10^{-4}$ \\
128&-& 2.1 $\times 10^{-4}$ \\
\end{tabular}
\end{table}

\end{document}